\documentclass{iopart}
\usepackage{iopams}
\usepackage[dvips]{graphicx}
\begin{document}
\title{ Strange matter in rotating compact stars}
\author{Sarmistha Banik$^{\rm (a)}$, Matthias Hanauske$^{\rm (b)}$ and Debades 
Bandyopadhyay$^{\rm (a)}$}
\address{$^{\rm (a)}$Saha Institute of Nuclear Physics, 1/AF Bidhannagar, 
Calcutta 700 064, India}
\address{$^{\rm (b)}$Institut f\"ur Theoretische Physik, J. W. Goethe 
Universit\"at, D-60054 Frankfurt am Main, Germany}

\begin{abstract}
We have constructed equations of state involving various exotic forms of matter 
with large strangeness fraction such as hyperon matter, Bose-Einstein 
condensates of antikaons and strange quark matter. First order phase 
transitions from hadronic to antikaon condensed and quark matter are considered
here. The hadronic phase is described by the relativistic field theoretical 
model. Later those equations of state are exploited to investigate models of 
uniformly rotating compact stars. The effect of rotation on
the third family branch for the equation of state involving only antikaon 
condensates is investigated. We also discuss the back bending phenomenon due 
to a first order phase transition from $K^-$ condensed to quark matter.
\end{abstract} 

\section{Introduction}
Neutron stars are born in type II supernova explosions of massive stars 
\cite{Bet}.
The matter in the neutron star interior is highly dense and cold. The matter
density could exceed by a few times normal nuclear matter density. Exotic
components of matter such as hyperons, antikaon condensates and quarks may 
appear there. Such a dense and cold matter can not be found in laboratories.
Therefore, neutron stars are very useful laboratories to study the properties
of dense matter \cite{Glenb,Web}.
  
It is a challenging problem to constrain the equation of state (EoS) of dense
matter in neutron stars. The theoretical investigation of mass-radius (M-R)
relationship of compact stars is important because this can be directly 
compared with observations. Consequently, the composition and EoS of dense 
matter may be probed. Though accurate masses have been measured for several
neutron stars in radio pulsar binaries, we do not have enough informations 
about their radii. Observations of thermal x-ray and optical radiation from
isolated neutron stars such as RX J185635-3754, may provide informations about
their radii \cite{Wal}. Another interesting candidate for radius 
measurement is x-ray bursters. Cottam et al. \cite{Cot} estimated
the gravitational redshift $z=$ 0.35 of three spectral lines in x-ray bursts
from EXO 0748-676. Recently, Villarreal and Strohmayer detected a 45 Hz 
rotational frequency for the same star \cite{Vil}. These observations lead to
a radius of 9.5-15 km and mass 1.5-2.3$M_{solar}$ for EXO 0748-676 \cite{Vil}. 
It is now being argued that the measurement of moment inertia of star A in 
double pulsar system PSR J0737-3039 where masses have been determined already,
could give an accurate estimate of the radius and put important constrain on 
the properties of neutron star matter \cite{Shap}.

Here we discuss various exotic forms of matter and their influence on equations
of state and structures of static and rotating compact stars. In this context,
we also report a new class of superdense stars called the third family
in static \cite{Ket,Schr,Bani1,Bani2,Han,Bani3} as well as rotating 
configurations \cite{Bani4} and the back bending phenomenon 
\cite{Glenb,Web,Bani4,Zud} which could be a signature for
a first order phase transition in the neutron star interior. In section 2, we
describe the calculation of EoS and structure of compact stars. Results are 
discussed in section 3. And the outlook is presented in section 4.

\section{Model}
Here we describe models of static and rotating compact stars. Such models are
constructed by solving Einstein's field equations,
\begin{equation}
{G^{\mu\nu} = R^{\mu\nu} - {1\over 2} g^{\mu\nu} R 
= 8{\pi}T^{\mu\nu}},
\end{equation}
where $R^{\mu\nu}$ is the Ricci tensor, $g^{\mu\nu}$ is the space-time metric,
$R$ is the scalar curvature and $T^{\mu\nu}$ is the energy-momentum tensor. 
The EoS is encoded in the energy-momentum tensor. The equilibrium structures
of non-rotating stars are calculated from Tolman-Oppenheimer-Volkoff (TOV)
equations. Here we are interested in the equilibrium structures of rotating 
compact stars. In this case, the metric has the form \cite{Cook,Ster}
\begin{equation}
ds^2 = - e^{\gamma + \rho} dt^2 + e^{2\alpha} (dr^2 + r^2 d{\theta}^2)
       + e^{\gamma - \rho} r^2 sin^2{\theta}(d\phi - \omega dt)^2 ~,
\end{equation}
where metric functions depend on $r$ and $\theta$. Angular velocity of local 
inertial frames is denoted by $\omega$.

In the following paragraphs, we
discuss two equations of state for dense matter. In the first case (case I), we 
consider a first order phase transition from hadronic matter including hyperons
to $K^-$ condensed matter followed by a second order
$K^0$ condensation. In the other case (case II), we have  first order phase
transitions from nuclear to $K^-$ condensed matter and then to quark matter.
  
For case I, we have hadronic phase, antikaon condensed phase and a mixed of 
those two phases. The constituents of matter in the hadronic phase are all the 
species of the baryon octet, electrons and muons.
Baryon-baryon interaction is mediated by the exchange of scalar and vector 
mesons. Here we adopt a relativistic field theoretical model to describe 
hadronic and 
antikaon condensed phase. The model is also extended to include hyperon-hyperon
interaction through two additional strangeness meson $f_0$(975) (hereafter
denoted as $\sigma^*$) and $\phi$(1020). Therefore the Lagrangian
density for baryon-baryon interaction is \cite{Bani1,Bani2,Sch96}       
\begin{eqnarray}
{\cal L}_B &=& \sum_B \bar\psi_{B}\left(i\gamma_\mu{\partial^\mu} - m_B
+ g_{\sigma B} \sigma - g_{\omega B} \gamma_\mu \omega^\mu
- g_{\rho B}
\gamma_\mu{\mbox{\boldmath t}}_B \cdot
{\mbox{\boldmath $\rho$}}^\mu \right)\psi_B\nonumber\\
&& + \sum_B \bar\psi_{B}\left(
g_{\sigma^* B} \sigma^* - g_{\phi B} \gamma_\mu \phi^\mu
\right)\psi_B\nonumber\\
&& + \frac{1}{2}\left( \partial_\mu \sigma\partial^\mu \sigma
- m_\sigma^2 \sigma^2\right) - U(\sigma) 
+ \frac{1}{2}\left( \partial_\mu \sigma^*\partial^\mu \sigma^*
- m_{\sigma^*}^2 \sigma^{*2}\right) \nonumber\\
&& -\frac{1}{4} \omega_{\mu\nu}\omega^{\mu\nu}
+\frac{1}{2}m_\omega^2 \omega_\mu \omega^\mu
- \frac{1}{4}{\mbox {\boldmath $\rho$}}_{\mu\nu} \cdot
{\mbox {\boldmath $\rho$}}^{\mu\nu}
+ \frac{1}{2}m_\rho^2 {\mbox {\boldmath $\rho$}}_\mu \cdot
{\mbox {\boldmath $\rho$}}^\mu  \nonumber\\
&& -\frac{1}{4} \phi_{\mu\nu}\phi^{\mu\nu}
+\frac{1}{2}m_\phi^2 \phi_\mu \phi^\mu~.
\end{eqnarray}
The scalar self-interaction term is
\begin{equation}
U(\sigma) = \frac{1}{3} g_2 \sigma^3 + \frac{1}{4} g_3 \sigma^4 ~.
\end{equation}

The antikaon condensed phase is composed of baryons
of the octet embedded in the condensate, electrons and muons.  
Similarly, we adopt a relativistic field theoretical approach for the 
description of (anti)kaon-baryon interaction. Here (anti)kaon-baryon 
interaction is treated in the same footing as baryon-baryon interaction.
The Lagrangian density for (anti)kaons
in the minimal coupling scheme is \cite{Bani1,Bani2,Gle99} 
\begin{equation}
{\cal L}_K = D^*_\mu{\bar K} D^\mu K - m_K^{* 2} {\bar K} K ~,
\end{equation}
where the covariant derivative
$D_\mu = \partial_\mu + ig_{\omega K}{\omega_\mu} + ig_{\phi K}{\phi_\mu}
+ i g_{\rho K}
{\mbox{\boldmath t}}_K \cdot {\mbox{\boldmath $\rho$}}_\mu$.
The isospin doublet for kaons
is denoted by $K\equiv (K^+, K^0)$ and that for antikaons is
$\bar K\equiv (K^-, \bar K^0)$. The effective mass of (anti)kaons in this
minimal coupling scheme is given by
\begin{equation}
m_K^* = m_K - g_{\sigma K} \sigma - g_{\sigma^* K} \sigma^* ~,
\end{equation}
In the mean field approximation adopted here, the meson fields are replaced by 
their
expectation values. The mean meson fields are denoted by $\sigma$, $\sigma^*$, 
$\omega_0$, $\phi_0$ and $\rho_{03}$. Each meson field equation in the antikaon
condensed phase has contributions from antikaon condensates \cite{Bani1,Bani2}.
These equations differ from those of pure hadronic phase by the antikaon 
condensate terms. 

The in-medium energies of antikaon for $s$-wave ($\bf k=0$) is 
\begin{equation}
\omega_{K^-,\: \bar K^0} = m_K^* - g_{\omega K} \omega_0 - g_{\phi K} \phi_0
\mp \frac{1}{2} g_{\rho K} \rho_{03} ~,
\end{equation}
where the isospin projection $I_{3\bar K} =\mp 1/2$ for $K^-$ meson and 
$\bar K^0$ meson respectively. 
The onsets of antikaon condensations are given by \cite{Bani1,Bani2},
\begin{eqnarray}
\mu_{K^-} = \mu_e ~, \\
\mu_{\bar K^0} &=& 0 ~,
\end{eqnarray}
where $\mu_{K^-}$ and $\mu_{\bar K^0}$ are respectively the chemical
potentials of $K^-$ and $\bar K^0$. 

The energy density ($\epsilon^h$) and pressure 
($P^h$) in hadronic phase are related by $P^h = \sum_i{\mu_i n_i} - \epsilon^h$.
Similarly, the pressure in the antikaon condensed phase follows from the 
relation 
$P^{\bar K} = \sum_i{\mu_i n_i} - \epsilon^{\bar K}$, where $n_i$ is the 
number density of i-th species. The expressions for $\epsilon^h$ and 
$\epsilon^{\bar K}$ are given as in Ref. \cite{Bani2}.

The mixed phase of antikaon condensed matter and hadronic matter is governed
by Gibbs phase rules and global conservation laws \cite{Glenb}. The Gibbs phase 
rules read, $P^h = P^{\bar K}$ and $\mu_B^h = \mu_B^{\bar K}$ where $\mu_B^h$
and $\mu_B^{\bar K}$ are chemical potentials of baryon B in hadronic and $K^-$
condensed matter respectively. The conditions for global charge neutrality and 
baryon number conservation are $(1-\chi) Q^h + \chi Q^{\bar K}=0$ and
$n_B = (1-\chi) n_B^h + \chi n_B^{\bar K}$ where $\chi$ is the volume fraction
in the condensed phase. 
Charges in hadronic and $K^-$ condensed phase are 
$Q^h = \sum_B q_B n^h_B -n_e -n_\mu$ and 
$Q^{\bar K}=\sum_B q_B n_B^{\bar K} -n_{K^-} - n_e - n_\mu$ respectively,
where $n_B^h$ and $n_B^{\bar K}$ are the number density of baryon B in pure 
hadronic and antikaon condensed phase and $n_{K^-}$, $n_e$ and $ n_\mu$ are 
number densities of antikaons, electrons and muons respectively.
The total energy density in the mixed phase is given 
by $\epsilon = (1-\chi) \epsilon^h + \chi \epsilon^{\bar K}$.
  
For case II, the first order phase transition from nuclear matter to $K^-$ 
condensed matter is treated in the same prescription as it is described above.
However, the hadronic and antikaon condensed phases are composed of neutrons,
protons, electrons and muons. In this case, we also consider a first order
phase transition from antikaon condensed matter to quark matter. 
The pure quark matter is composed of $u$, $d$ and $s$ quarks. We describe the
EoS of pure quark matter in the MIT bag model \cite{Far}. The free energy 
density of non-interacting quarks is given by,
\begin{equation}
\Omega = \frac {3}{\pi^2} \sum_{i=u,d,s} \int_0^{p_{F,i}} dp p^2 
(\sqrt {p^2 + m_i^2} - \mu_i) + B,~ 
\end{equation}
where B is the bag energy density. The pressure in the quark phase 
is given by $P^Q=-\Omega$ and the energy density ($\epsilon^Q$) is obtained 
from the Gibbs-Duhem relation. And the mixed phase is governed by the Gibbs 
phase rules and conservation laws. 

\begin{figure}[t]
\begin{center}
\includegraphics[height=8cm]{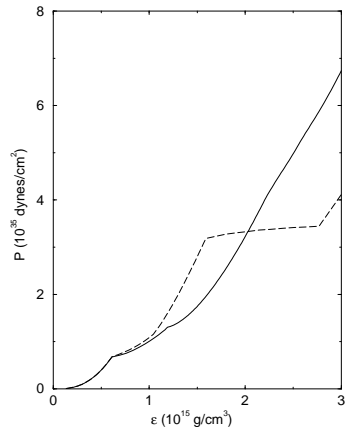} 
\caption{The equation of state, pressure $P$ versus energy density
$\varepsilon$, for  two cases as described in the text.}
\end{center}
\end{figure}

\section{Results \& Discussions}
In this calculation, we adopt GM1 parameter set \cite{Gle2} where nucleon-meson
coupling constants are determined from the nuclear matter saturation 
properties. The vector meson coupling constants for 
(anti)kaons and hyperons are determined from the quark model \cite{Sch96} 
whereas the scalar meson coupling constants for hyperons and antikaons are 
obtained from the potential depths of  hyperons and antikaons in normal nuclear 
matter \cite{Bani2,Sch96}. As the phenomenological fit to the $K^-$ atomic 
data yielded a very strong real part of antikaon potential 
$U_{\bar K} = -180 \pm 20$ MeV \cite{Fri}, we 
perform this calculation with an antikaon optical potential of -160 MeV at 
normal nuclear matter density ($n_0=0.153 fm^{-3}$). The coupling 
constants for strange mesons with hyperons and (anti)kaons are taken from
Ref.\cite{Bani2,Sch96}. Also, we consider a bag constant $B^{1/4}$ = 200 MeV
and strange quark mass $m_s$ = 150 MeV for quark EoS in case II.

The equations of state, pressure versus energy density, for case I and case II
are exhibited in Figure 1 by solid and dashed lines respectively. 
For both cases, the upper and lower boundaries of the mixed phase of hadronic 
and $K^-$ condensed matter are identified by two kinks. The mixed phase 
begins at energy density $\epsilon =$ 6.11$\times 10^{14}$ g/cm$^3$ 
(or 2.23$n_0$). For case I, $\Lambda$ hyperons are populated in the
mixed phase at 2.51$n_0$ whereas $\Xi^-$ and $\Sigma^-$ hyperons are
populated at much higher densities. A second order $K^0$ condensation occurs
in case I just after the mixed phase is over at 
$\epsilon =$ 1.17$\times 10^{15}$ g/cm$^3$. For case II, the phase transition
to quark matter is delayed due to the early onset of $K^-$ condensation
depending on the parameters. The first order $K^-$ condensed-quark matter phase 
transition begins at $\epsilon =$ 1.59$\times 10^{15}$ g/cm$^3$ and ends at 
$\epsilon =$ 2.77$\times 10^{15}$ g/cm$^3$. In this case, the extent of the 
mixed phase is quite large. We note that the overall EoS for case I is softer
than that of case II. 

\begin{figure}[t]
\begin{center}
\includegraphics[height=8cm]{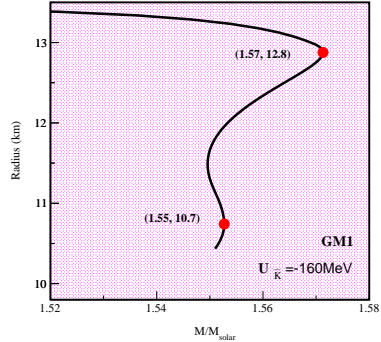} 
\caption{Mass-radius relationship of non-rotating stars calculated with the 
equation of state for case I.}
\end{center}
\end{figure}
In Figure 2, the mass-radius relationship of non-rotating stars is displayed 
for the EoS of case I. The filled circles correspond to the maximum masses.
We obtain a maximum neutron star mass of 1.57$M_{solar}$ corresponding to a
radius of 12.8 km as shown in the figure. It is found that after the positive
slope neutron star branch, there is an unstable region followed by another 
positive slope compact star branch. This new branch of compact stars is
the result of the discontinuity in the speed of sound due to the kinks or first
order phase transition in the EoS for case I. This new sequence of superdense 
stars beyond the neutron star branch is called the third family branch. From 
the study of fundamental mode of radial vibration, it was found
that the third family branch was a stable one \cite{Bani2}. The compact stars 
in the third family branch have different compositions and smaller radii than
their counterparts on the neutron star branch \cite{Bani2,Bani3}. The maximum
mass of the compact star in the third family branch is 1.55$M_{solar}$ 
corresponding to a radius 10.7 km. 

\begin{figure}[t]
\begin{center}
\includegraphics[height=8cm,width=7cm,height=9cm]{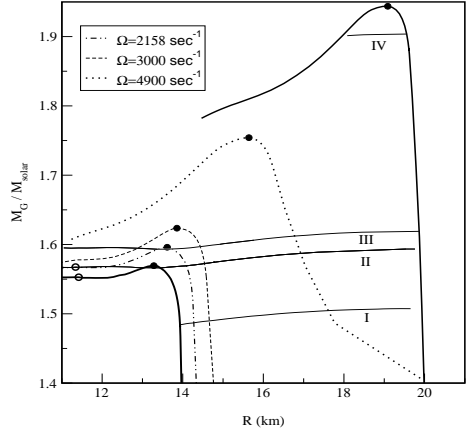} 
\caption{Gravitational mass as a function of equatorial radius for case I.
The static and mass shedding limit sequences along with fixed angular velocity
sequences are displayed here. Also, normal and supramassive sequences are
plotted.}
\end{center}
\end{figure}
Now we discuss the mass-equatorial radius relationship of rotating sequences
calculated with the EoS of case I. This is shown in Figure 3. In the preceding
paragraph, we have discussed the third family of compact stars in the static 
sequence calculated with this EoS. It is interesting to watch what happens to 
this third family branch in rotating cases. In Figure 3, the extreme left and
right curves denote the static and mass shedding limit sequences. Each compact 
star on the mass shedding limit rotates with a maximum frequency called
the Kepler frequency and it becomes unstable beyond this frequency.
Fixed angular velocity curves are also shown in this figure. On each curve, the
maximum mass in the neutron star branch is denoted by solid circles whereas the 
maximum mass in the third family branch is represented by open circles. We
observe that with increasing rotation, the maximum star masses also increase and
are shifted to smaller central energy densities or larger radii. This is
attributed to the presence of centrifugal force in rotating stars. On the other
hand, we note that the third family of compact stars disappears beyond the
fixed angular velocity curve $\Omega$ = 2158 sec$^{-1}$. The reason is
the strengthening of the centrifugal force with increasing rotation. 
Consequently, fast rotating stars could not probe the high density part of the 
EoS which is responsible for the origin of a third family branch.   

In Figure 3, we also plot fixed baryon number evolutionary sequences. Here two
normal sequences are shown by curves I and II and two supramassive sequences
by curves III and IV. Fixed baryon number sequences provide important
informations about isolated rotating stars as they slow down with time due to
the loss of energy and angular momentum through electromagnetic and 
gravitational radiation. We observe from the figure that non-rotating stars are
the members of normal sequences. However, no member of supramassive sequences
is a non-rotating compact star. When a supramassive star loses sufficient
angular momentum such that the centrifugal support against
self gravity is impossible, it would then collapse to a black hole. This was
considered as a possible scenario for gamma ray bursts \cite{Vie}. 

\begin{figure}[t]
\begin{center}
\includegraphics[height=8cm]{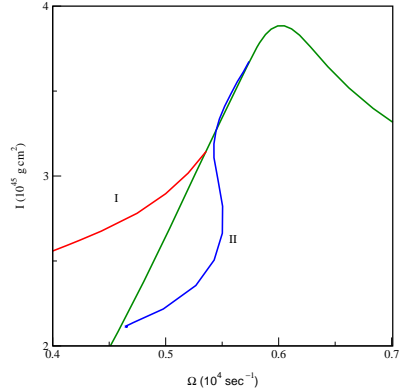} 
\caption{Moment of inertia (I) is plotted with angular velocity ($\Omega$) for
the EoS of case II. Normal and supramassive sequences are shown by curves I 
and II respectively. The other curve is the mass shedding limit sequence.}
\end{center}
\end{figure}

We also investigate the mass-equatorial radius relationship of the static, 
mass shedding limit and fixed angular velocity sequences calculated with the 
EoS for case II. However, we do not find any third family solution in the
static and rotating cases. But we find some interesting results from the
study of a supramassive sequence. We display moment of inertia (I) with
angular velocity ($\Omega$) in Figure 4. Curve I implies the normal sequence 
with baryon number $N_b = (2.15 \times 10^{57})$ whereas curve II is a 
supramassive sequence with $N_b = (2.45 \times 10^{57})$. The other curve is
the mass shedding limit sequence. For the normal sequence, the moment of inertia
always decreases with decreasing angular velocity. However, curve II exhibits  
that after the initial spin down of the neutron star along this sequence, 
there is a spin up followed by another spin down. This is known 
as the back bending phenomenon. We observe that this back bending phenomenon is 
the result of the first order phase transition from antikaon condensed to
quark matter. It is worth mentioning here that the supramassive neutron star 
is stable on the back bending segment \cite{Bani4}. We also note that the
period of the neutron star in the back bending part varies from 1.10 ms to 
1.35 ms corresponding to the variation in angular velocity from 5700 s$^{-1}$
to 4660 s$^{-1}$. The minimum observed pulsar period is so far 1.56 ms. Future
pulsar surveys would tell us whether pulsars rotating faster than the currently
known one exist in nature or not.
\section{Outlook}
We have investigated equations of state including exotic forms of matter and 
computed the structures of static and rotating compact stars. Many interesting
results such as the third family of compact stars containing exotic matter,
the back bending effect which could signal a first order phase transition in 
the neutron star interior, are obtained here. Our results are sensitive to
input parameters such as antikaon optical potential depth and bag constant.
Future experiments on deeply bound states of antibaryon-nucleus at GSI, 
Germany and $\bar K$-nucleus at Japan Proton Accelerator Complex
(J-PARC) will provide opportunity to investigate dense and cold matter in 
laboratories and determine those parameters accurately. It would be possible 
to constrain the composition and EoS of dense and cold matter in compact stars 
from observations and experiments in future.

\noindent{\bf Acknowledgments:}

\noindent This work is supported by the Department of Science and Technology 
(DST), Government of India, German Academic Exchange Service (DAAD), Germany, 
Department of Science of Technology, Government of South Africa and 
GSI, Germany.

\end{document}